\documentclass[a4paper,fleqn,usenatbib]{mnras}

\usepackage{newtxtext,newtxmath}

\usepackage[T1]{fontenc}
\usepackage{ae,aecompl}

\usepackage{graphicx}	
\usepackage{amsmath}	
\usepackage{amssymb}	

\title[Optical timing of PSR J1023+0038 with Aqueye+]{Precise optical timing of PSR J1023+0038, the first millisecond pulsar detected with Aqueye+ in Asiago}

\author[L. Zampieri, A. Burtovoi, M. Fiori et al.]{
Luca Zampieri,$^{1}$\thanks{E-mail: luca.zampieri@inaf.it}
Aleksandr Burtovoi,$^{2,1}$
Michele Fiori,$^{3,1}$
Giampiero Naletto,$^{3,1}$
\newauthor
Alessia Spolon,$^{3,1}$
Cesare Barbieri,$^{3,1}$
Alessandro Papitto$^{4}$
and Filippo Ambrosino$^{5}$
\\
$^{1}$INAF - Osservatorio Astronomico di Padova, Vicolo dell'Osservatorio 5, 35122, Padova, Italy\\
$^{2}$Centre of Studies and Activities for Space (CISAS) ``G. Colombo'', University of Padova, Via Venezia 15, 35131 Padova, Italy\\
$^{3}$Department of Physics and Astronomy, University of Padova, Via F. Marzolo 8, 35131, Padova, Italy\\
$^{4}$INAF - Osservatorio Astronomico di Roma, Via Frascati 33, 00076, Monteporzio Catone (RM), Italy\\
$^{5}$INAF - Istituto di Astrofisica e Planetologia Spaziali, Via del Fosso del Cavaliere 100, 00133 Rome, Italy
}

\date{Accepted XXX. Received YYY; in original form ZZZ}

\pubyear{2019}

\begin{document}
\label{firstpage}
\pagerange{\pageref{firstpage}--\pageref{lastpage}}
\maketitle

\begin{abstract}
We report the first detection of an optical millisecond pulsar with the fast photon counter Aqueye+ in Asiago.
This is an independent confirmation of the detection of millisecond pulsations from PSR J1023+0038 obtained with SiFAP at the Telescopio Nazionale Galileo.
We observed the transitional millisecond pulsar PSR J1023+0038 with Aqueye+ mounted at the Copernicus telescope in January 2018. Highly significant pulsations were detected. The rotational period is in agreement with the value extrapolated from the X-ray ephemeris, while the time of passage at the ascending node is shifted by $11.55 \pm 0.08$ s from the value predicted using the orbital period from the X-rays. An independent optical timing solution is derived over a baseline of a few days, that has an accuracy of $\sim 0.007$ in pulse phase ($\sim 12$ $\mu$s in time). This level of precision is needed to derive an accurate coherent timing solution for the pulsar and to search for possible phase shifts between the optical and X-ray pulses using future simultaneous X-ray and optical observations.
\end{abstract}




\section{Introduction}
\label{sect:introduction}


Millisecond radio pulsars were discovered in 1982 \citep{1982Natur.300..728A,1982Natur.300..615B}. The first confirmation of the long sought evolutionary link between these sources and accreting Low Mass X-ray Binaries (LMXBs) came in 1998 with the discovery of pulsations at 2.5 ms in SAX J1808.4-3658 \citep{1998Natur.394..344W}. This showed that old, low magnetic field pulsars are re-accelerated (``recycled'') through mass and angular momentum transfer from a companion star in a previous LMXB phase.

However, it was only quite recently that direct evidence that some of these systems can swing between a rotation-powered millisecond pulsar phase and an accretion phase was gathered, thanks to the amazing discovery of an ordinary radio millisecond pulsar switching off and turning on as an accretion-powered, X-ray millisecond pulsar (PSR J1824-2452;  \citealt{2013Natur.501..517P}). At present, we know other two systems that behave in a similar way: PSR J1023+0038 \citep{2009Sci...324.1411A} and PSR J1227-4853 \citep{2010A&A...515A..25D}. These fast spinning, weakly magnetized neutron stars (NSs) are called transitional millisecond pulsars (tMSPs) and typically have low mass companion stars ($\sim 0.5 M_\odot$).

\begin{table}
\scriptsize
\centering
\caption{Log of the 2018 observations of PSR J1023+0038 taken with Aqueye+ at the 1.8 m Copernicus telescope in Asiago.}
\label{tab:observations}
\begin{tabular}{rrrr}
\hline
\hline
\multicolumn{1}{c}{Observation ID} & \multicolumn{1}{c}{Start time} & \multicolumn{1}{c}{Start time} & \multicolumn{1}{c}{Duration} \\
 & \multicolumn{1}{c}{(UTC)} & \multicolumn{1}{c}{(MJD)} & \multicolumn{1}{c}{(s)}	
\\
\hline
 20180122-011224  &  Jan 22$^{\rm a}$  00:20:10.2  &  58140.014007   &  899.4  \\   
 20180122-012903  &          00:36:48.3  &  58140.025559   &  899.4  \\
 20180122-014725  &          00:55:10.3  &  58140.038314   &  899.4  \\
 20180122-020320  &          01:11:05.4  &  58140.049368   &  899.4  \\
 20180122-022214  &          01:29:59.5  &  58140.062494   &  899.4  \\
 20180122-023844  &          01:46:29.5  &  58140.073953   &  899.4  \\
 20180122-030122  &          02:09:07.6  &  58140.089671   &  1799.4  \\   
 20180122-034854  &          02:56:39.7  &  58140.122682   &  1799.2  \\   
 20180122-042617  &          03:34:02.9  &  58140.148645   &  1799.4  \\   
 20180122-050133  &          04:09:20.0  &  58140.173148   &  1799.4  \\   
 20180123-013839  &  Jan 23$^{\rm a}$  00:46:30.3  &  58141.032295   &  899.4  \\ 
 20180123-015619  &          01:04:10.4  &  58141.044565   &  899.4  \\ 
 20180123-021846  &          01:26:36.4  &  58141.060144   &  899.4  \\ 
 20180123-023447  &          01:42:38.5  &  58141.071279   &  899.4  \\ 
 20180123-025439  &          02:02:30.6  &  58141.085076   &  1799.4  \\ 
 20180123-033110  &          02:39:01.7  &  58141.110436   &  1799.4  \\ 
 20180123-040447  &          03:12:37.8  &  58141.133771   &  1799.3  \\ 
 20180123-044543  &          03:53:34.0  &  58141.162199   &  1799.4  \\ 
 20180124-023319  &  Jan 24$^{\rm a}$  01:41:14.4  &  58142.070306   &  1799.4  \\ 
 20180124-031307  &          02:21:02.5  &  58142.097946   &  1799.4  \\ 
 20180124-034843  &          02:56:39.6  &  58142.122681   &  1799.4  \\ 
 20180124-042332  &          03:31:27.8  &  58142.146849   &  1799.4  \\ 
 20180124-045838  &          04:06:33.9  &  58142.171226   &  1799.4  \\ 
 20180124-053343  &          04:41:39.0  &  58142.195590   &  1799.4  \\ 
 20180125-014934  &  Jan 25$^{\rm a}$  00:57:35.0  &  58143.039989   &  1799.4  \\ 
 20180125-024737  &          01:55:38.2  &  58143.080303   &  1799.4  \\ 
 20180125-034559  &          02:54:00.4  &  58143.120838   &  1799.4  \\ 
 20180125-042058  &          03:28:58.5  &  58143.145122   &  1799.4  \\ 
 20180125-045849  &          04:06:50.6  &  58143.171419   &  1799.4  \\ 
 20180125-054140  &          04:49:40.7  &  58143.201166   &  899.4  \\ 
\hline
\end{tabular}
\begin{minipage}{8.2 cm}
Start times refer to the Solar system barycenter.\\
$^{\rm a}$Seeing: Jan 22 $\sim 2.9$'', Jan 23 $\sim 2.9$'', Jan 24 $\sim 2.0$'', Jan 25 $\sim 2.3$''.
\end{minipage}
\end{table} 

PSR J1023+0038, located at a distance of $1.37 \pm 0.04$ kpc \citep{2012ApJ...756L..25D}, was initially classified as a Cataclysmic Variable \citep{2002PASP..114.1359B}.
Subsequent observations carried out with the Green Bank Telescope showed that in 2007 the source was, in fact, a radio pulsar with a rotational period of 1.69 ms orbiting a $\sim 0.2 M_\odot$ companion with a period of 4.75 hours \citep{2009Sci...324.1411A}. It was then realized that, between 2000 and 2001, the source had an accretion disc, which subsequently disappeared following the appearance of the radio millisecond pulsar. In June 2013 an inverse transition took place, with the disappearance of the radio pulses and the reappearance of the accretion disc, with a strong double-peaked H$\alpha$ emission observed in the optical spectrum \citep{2013ATel.5514....1H,2014ApJ...781L...3P,2014ApJ...790...39S}. In the disc state the average optical and X-ray fluxes increase by a factor of $\sim 2-3$ \citep{2014MNRAS.444.1783C} and $\sim 10$ \citep{2014ApJ...790...39S}, respectively, the GeV $\gamma$-ray flux is 3-5 times higher \citep{2017ApJ...836...68T}, while only an upper limit exists on the TeV $\gamma$-ray flux \citep{2016ApJ...831..193A}. As for the radio emission, bright continuum emission with a flat spectrum is observed, with radio flares occurring at certain phases \citep{2015ApJ...809...13D,2018ApJ...856...54B}.

During the accretion phase significant X-ray variability is observed in PSR J1023+0038 on time scales of tens of seconds. {\it XMM-Newton} observations show a puzzling trimodal behaviour in the 0.3-10 keV band. PSR J1023+0038 spends about 70-80\% of the time in a stable {\it high mode} with a X-ray luminosity $L_X \sim 7 \times 10^{33}$ erg s$^{-1}$. The source then unpredictably switches to a {\it low mode} with a much lower luminosity ($L_X \sim 10^{33}$ erg s$^{-1}$;  \citealt{2015ApJ...806..148B,2016A&A...594A..31C,2018A&A...611A..14C}), often correlated with enhancements of the radio emission \citep{2018ApJ...856...54B}. Sporadic X-ray flaring episodes are also observed reaching luminosities $L_X \sim 10^{34}$ erg s$^{-1}$ ({\it flaring mode}). Coherent X-ray pulsations at the NS spin period are detected only when the system is in the {\it high mode} \citep{2015ApJ...807...62A}. The luminosity in this mode is lower than that of a typical LMXB. Notably, the spin rate of the pulsar during the accretion state is close to the value measured during the radio pulsar state \citep{2016ApJ...830..122J}.

Optical and infrared observations of PSR J1023+0038 revealed significant variability and flaring activity \citep{2015MNRAS.453.3461S,2018MNRAS.477..566S,2018MNRAS.474.3297H,2018MNRAS.477.1120K,2018ApJ...858L..12P}, with an optical polarization degree up to $\sim 1$\% \citep{2016A&A...591A.101B}. A striking result coming from optical observations of PSR J1023+0038 was the discovery of millisecond optical pulsations with  the spin period of the pulsar, produced in a region only a few tens of km away from the NS \citep{2017NatAs...1..854A}, most likely caused by an active rotation-powered pulsar even in the disc state.

The origin of the millisecond pulsations and the multiwavelength variability of PSR J1023+0038 in the disc state is matter of lively debate at present. Several possibilities have been explored, from emission of a rotational powered pulsar \citep{2014ApJ...790...18T,2014MNRAS.444.1783C,2014ApJ...797..111L}, to a pulsar in the propeller stage \citep{2014MNRAS.438.2105P,2015ApJ...807...33P}, a pulsar accreting from a 'dead' disc \citep{2012MNRAS.420..416D}, or a pulsar switching between different regimes over short ($\sim$10 s) timescales \citep{2014ApJ...795...72L,2016A&A...594A..31C,2018A&A...611A..14C}.

A crucial aspect to consider in all these models is the existence of millisecond optical pulsations. In order to understand the properties of these pulsations and their relation to the X-ray pulsations, it is important to foster the present observational framework and to increase the number and accuracy of the available optical measurements.
In this context, we report the independent confirmation of the \cite{2017NatAs...1..854A} detection of millisecond pulsations from PSR J1023+0038 with the fast photon counter Aqueye+ mounted at the Copernicus telescope in Asiago.

The plan of the paper is the following. In Section~\ref{sect:observations} we report the Aqueye+ observations of PSR J1023+0038 carried out in January 2018, in Section~\ref{sect:results} we show the results of our timing analysis and in Section~\ref{sect:discussion} we shortly discuss their potential impact in future simultaneous mutiwavelength campaigns.


\section{Observations and data analysis}
\label{sect:observations}

We observed PSR J1023+0038 with Aqueye+ mounted at the Copernicus telescope in Asiago, Italy. Aqueye+\footnote{https://web.oapd.inaf.it/zampieri/aqueye-iqueye/index.html} is a fast photon counter with a field of view of few arcsec and the capability of time tagging the detected photons with sub-ns time accuracy \citep{2009JMOp...56..261B,2013SPIE.8875E..0DN,2015SPIE.9504E..0CZ}. A total of 30 acquisitions were performed between January 22 and 25, 2018, each lasting either $\sim 900$ s or $\sim 1800$ s. The overall on-source observing time is $\sim 44.1$ ks. The log of the observations is shown in Table~\ref{tab:observations}. The sky background was regularly monitored between on-target observations. The average background-subtracted rate of PSR J1023+0038 varied between $\sim 800$ and $\sim 2000$ c/s, mostly because of intrinsic source variability (flares).

The data reduction is performed with a dedicated software\footnote{QUEST v. 1.1.5, see \cite{2015SPIE.9504E..0CZ}.}. The whole acquisition and reduction chain ensures an absolute accuracy of $\sim 0.5$ ns relative to UTC \citep{2009A&A...508..531N}. The photon arrival times are barycentered using TEMPO2 in TDB time units \citep{2006MNRAS.372.1549E,2006MNRAS.369..655H}, using the position of \cite{2016ApJ...830..122J} (RA=10:23:47.687198, DEC=+00:38:40.84551 at MJD 54995) and the JPL ephemerides DE405. We corrected for the motion of the NS along the orbit using the orbital period $P_{orb} = 0.1980963155$ days and projected semi-major axis $a = 0.343356$ light-seconds of \cite{2016ApJ...830..122J}.


\begin{figure}
	\includegraphics[angle=0, width=0.95\columnwidth]{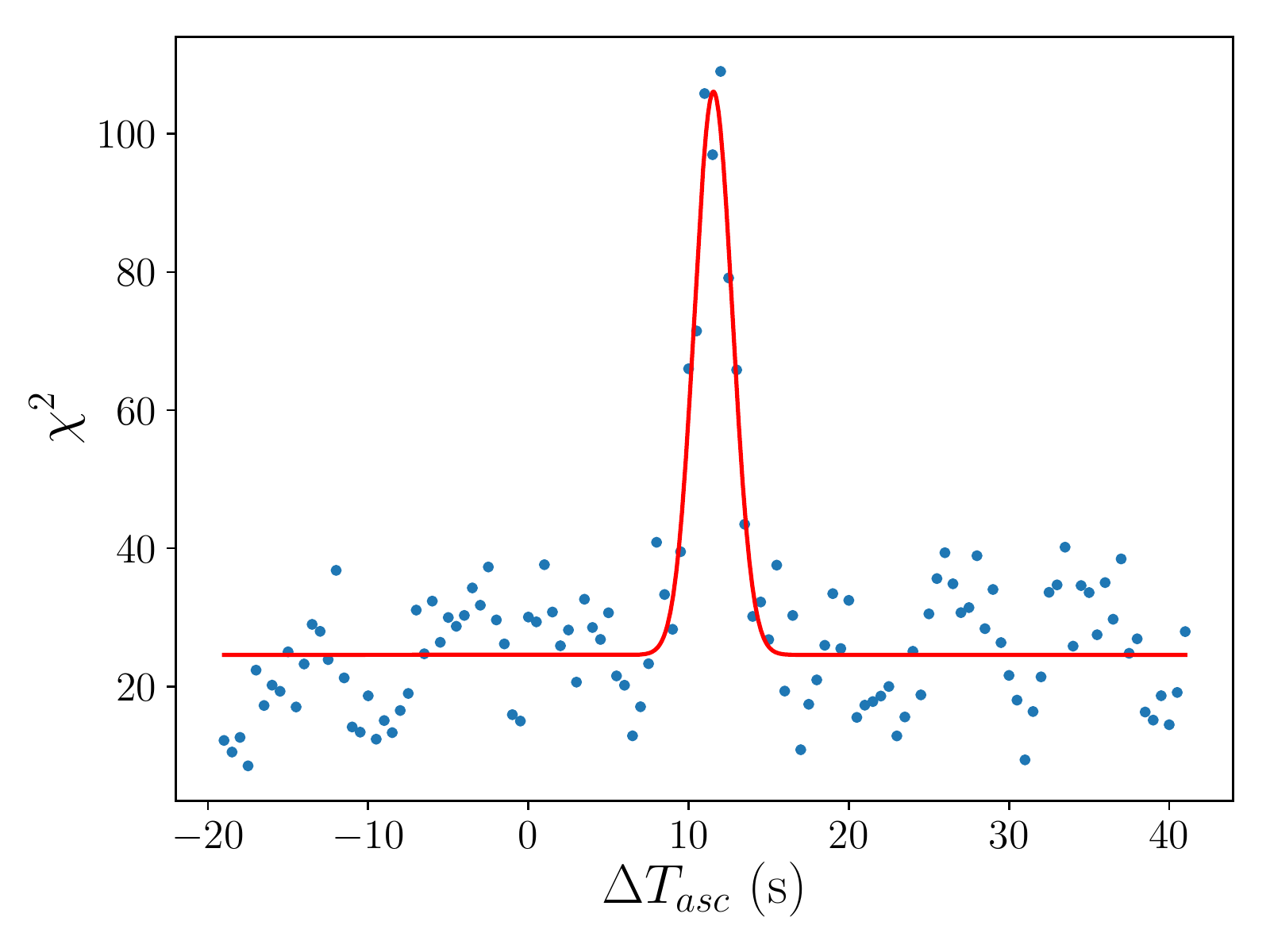}
    \caption{Epoch folding search of the time of passage at the ascending node for the January 25, 2018 Aqueye+ observation of PSR J1023+0038. Time is measured from the value of $T_{asc}$ predicted using the orbital period from the X-ray ephemeris \citep{2016ApJ...830..122J}.
}
    \label{fig:chi2}
\end{figure}

\section{Results}
\label{sect:results}

When PSR J1023+0038 transited to the accretion state, the time of passage of the pulsar at the ascending node $T_{asc}$ showed a $\sim 20-30$ s shift from the value measured during the radio pulsar phase. In addition to that, variations of a few seconds around an approximately sinusoidal modulation are observed \citep{2016ApJ...830..122J,papitto2019}. Thus, following \cite{2017NatAs...1..854A} we performed an accurate epoch folding search for the value of $T_{asc}$ for our epoch by folding the corrected event lists of January 25, 2018 with the spin period extrapolated from the X-ray ephemeris ($P_X = 1.6879874462$ ms at $T_0 = 58140.0$ MJD; \citealt{2016ApJ...830..122J}) and with 16 bins per period. The maximum of the $\chi^2$ is obtained for $T_{asc} = 58140.0915932 \pm 9 \times 10^{-7}$ MJD, shifted by $11.55 \pm 0.08$ s from that predicted using the orbital period from the X-ray ephemeris (see Fig.~\ref{fig:chi2}).
For this value of $T_{asc}$ the variance of the pulse profile with respect to a constant ($\chi^2=100$ for 15 degrees of freedom) gives a probability $<10^{-14}$ that the pulsation is caused by a statistical fluctuation, corresponding to a detection significance of $7.7 \sigma$.

Using the actual value of $T_{asc}$, we then performed a phase fitting of the Aqueye+ data, corrected for the binary motion, following the approach presented in previous works \citep{2012A&A...548A..47G,2014MNRAS.439.2813Z,2019MNRAS.482..175S}. We folded seperately the four nights of observations reported in Table~\ref{tab:observations} using as reference rotational period the value $P_{\rm init} = 1.687987440$ ms and 16 bins per period. We detected the two peaks of the pulse profile in each night. They were fitted with the sum of two harmonically-related sinusoids plus a constant \citep{2017NatAs...1..854A}. To improve the accuracy of the measurement, we performed the fit fixing the separation $\Delta\phi$ and ratio $\rho$ of the two sinusoids at the values obtained from fitting the pulse shape of the best observing night (Jan 25, 2018): $\Delta\phi = 0.419$, $\rho = 1.27$. The typical uncertainty on the phase measurement of each single night is $\sim 0.007$ (or 12 $\mu$s).

\begin{figure}
	\includegraphics[angle=0, width=0.95\columnwidth]{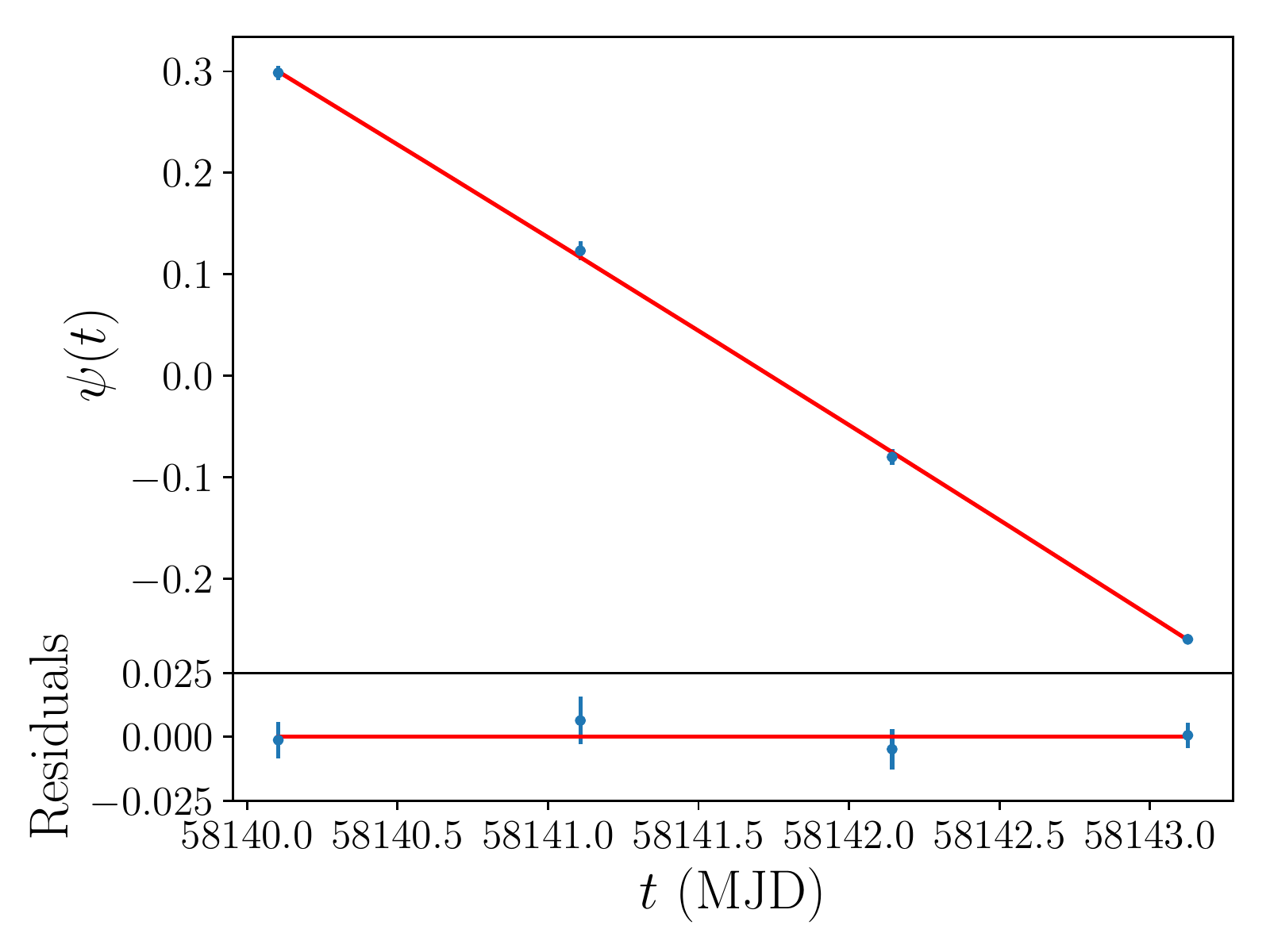}
    \caption{Phase drift $\psi(t)$ of PSR J1023+0038 with respect to a uniform rotation with period $P_{\rm init}$, fitted with a second-order polynomial ({\it solid} line). Time $t$ is expressed in MJD. Residuals of the fit are shown in the bottom panel.
}
    \label{fig:timingsol}
\end{figure}

\begin{table}
\caption{Timing solution of the January 2018 Aqueye+ observations of PSR J1023+0038.}
\label{tab:timingsol}
\centering
\begin{tabular}{l l}
\hline\hline
\multicolumn{2}{c}{Timing solution of PSR J1023+0038 in January 2018}	\\
\hline
$t_0$		& 58140 MJD \\
$\phi_0$	& $0.3186 \pm 0.0073$ 	\\
$\nu_0$		& $592.42146759  \pm 1.4 \times 10^{-7}$ Hz	\\
$|\dot \nu_0|$	& $< 9 \times 10^{-13}$ Hz$^2$ $^{\rm a}$ \\
\hline
\end{tabular}
\begin{minipage}{8.2 cm}
Uncertainties are calculated taking the square root of the diagonal terms in the covariance matrix of the fit.\\
$^{\rm a}$Upper limit estimated from the uncertainty on this parameter.
\end{minipage}
\end{table}

The measured phases show a drift $\psi(t)$\footnote{As defined in \cite{2014MNRAS.439.2813Z} and \cite{2019MNRAS.482..175S}.} with respect to a uniform rotation with period $P_{\rm init}$ (Fig.~\ref{fig:timingsol}). We fitted it with a second-order polynomial and obtained the timing solution reported in Table \ref{tab:timingsol}.
The rotational period ($P = 1/\nu_0 = 1.68798744596 \pm 3.9 \times 10^{-10}$ ms) 
is consistent within the uncertainties with the value extrapolated at $t_0 = 58140$ MJD from the X-ray ephemeris ($P_X = 1.687987446202 \pm 4 \times 10^{-12}$ ms; \citealt{2016ApJ...830..122J}). Only an upper limit to the first derivative of frequency could be obtained. Timing noise is present at a level of $\sim 0.01$ in phase. We note that additional terms accounting for the phase residuals induced by the orbital motion were not included in the fit because the reduced $\chi^2$ of the second-order polynomial fit is smaller than unity (0.9) and because of the limited number of phase measurements available.

\begin{figure}
	\includegraphics[angle=0, width=0.95\columnwidth]{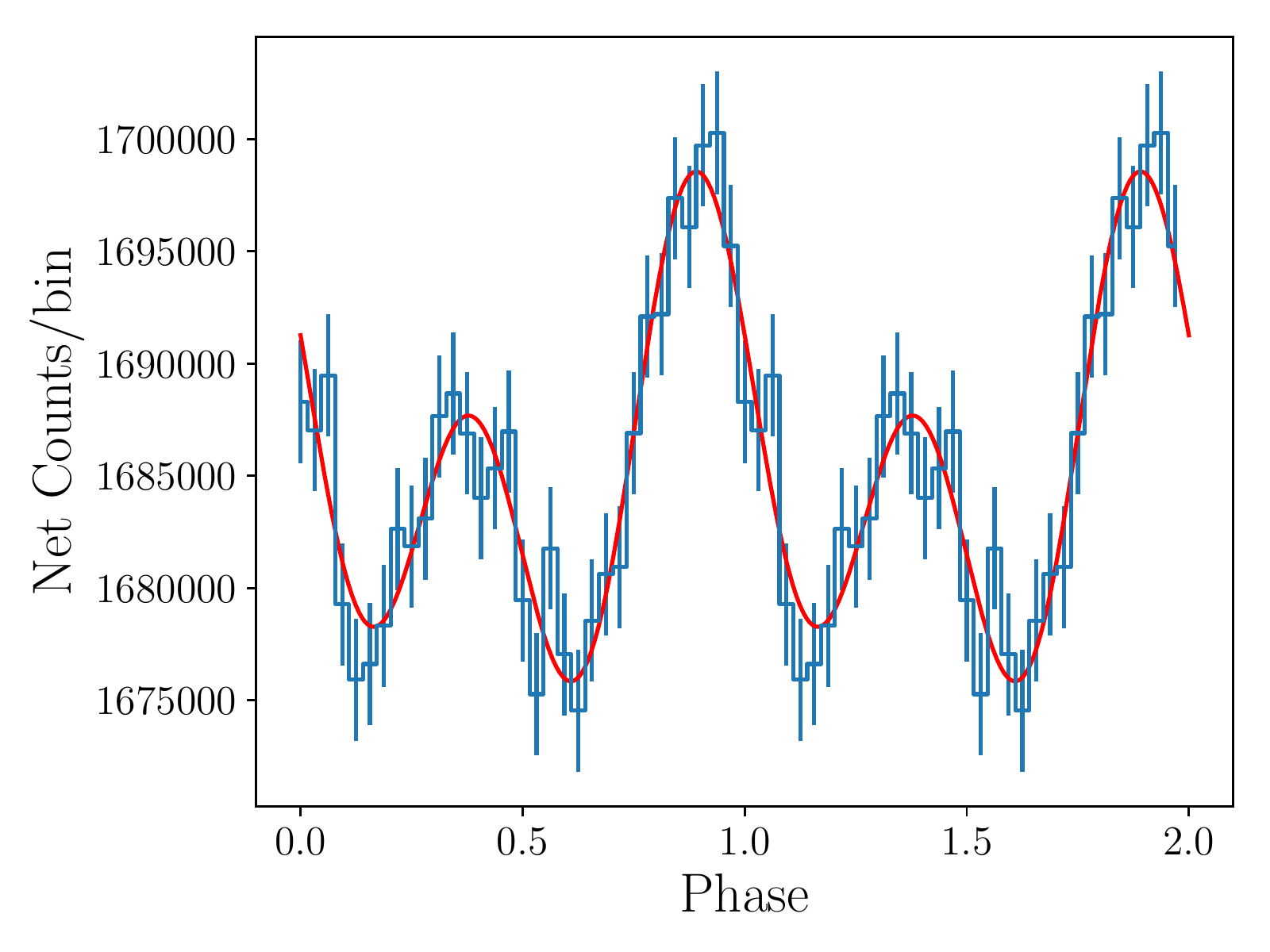}
    \caption{Background-substracted pulse shape of PSR J1023+0038 obtained folding all the Aqueye+ observations of January 2018 with the timing solution reported in Table~\ref{tab:timingsol} and with 32 bins per period. Two rotational phases are shown. The solid line shows the fit with two sinusoidal components plus a constant (model taken from \citealt{2017NatAs...1..854A}).}
    \label{fig:pulse}
\end{figure}

The final background-subtracted pulse shape obtained from folding all the January 2018 Aqueye+ observations with the timing solution reported in Table~\ref{tab:timingsol} and with 32 phase bins is shown in Fig.~\ref{fig:pulse}. A fit with the same model (two sinusoidal components plus a constant) adopted by \cite{2017NatAs...1..854A} and \cite{papitto2019} leads to consistent values of the parameters. The amplitudes of our two sinusoidal components are $A_1 = 0.0033 \pm 0.0002$ and $A_2 = 0.0046 \pm 0.0002$, and the phases are $\phi_1 = 0.671 \pm 0.012$ and $\phi_2 = 0.261 \pm 0.004$. The total fractional amplitude of the signal is $A = (A_1^2 + A_2^2)^{1/2} = 0.6$\%.
The variance of the pulse profile with respect to a constant ($\chi^2=440$ for 31 degrees of freedom) gives a probability $<10^{-73}$ that the pulsation is caused by a statistical fluctuation, corresponding to a detection significance of $18.2 \sigma$.

\begin{figure}
	\includegraphics[angle=0, width=0.9\columnwidth]{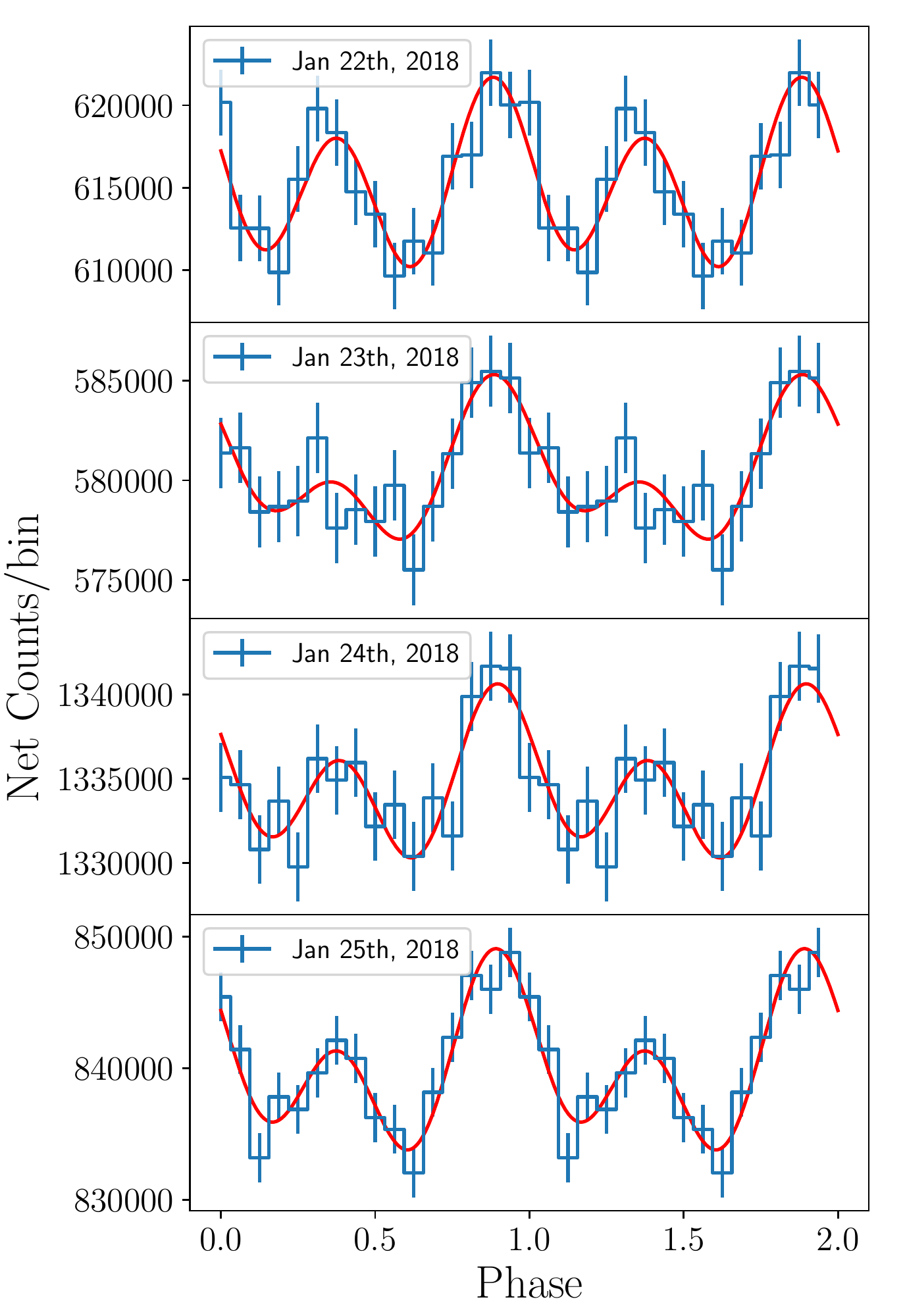}
    \caption{Background-substracted pulse shape of PSR J1023+0038 obtained folding each night of January 2018 with the timing solution reported in Table~\ref{tab:timingsol} and with 16 bins per period. Two rotational phases are shown. The solid lines show the fits with two sinusoidal components plus a constant (model taken from \citealt{2017NatAs...1..854A}), where we keep the relative phases of the two sinusoidal components fixed ($\Delta \phi = 0.41$).
}
    \label{fig:fourpulses}
\end{figure}

We used the same model to investigate possible night-to-night variations of the pulse shape, keeping the relative phases of the two sinusoidal components fixed at the value inferred from the previous fit ($\Delta \phi = 0.41$). 
The amplitudes of the two fitting sinusoids are in the interval $A_1=0.002-0.005$ and $A_2=0.003-0.007$.
The corresponding curves are shown in Fig. \ref{fig:fourpulses}. 
There are some differences in the quality of the pulse profiles in the various nights, but they do not appear to be caused by variations of the seeing conditions and/or depend in a straightforward way on the average source rate.
Indeed, the night with the best seeing and the highest background-subtracted average rate (Jan 24; see Table~\ref{tab:observations} and Fig. \ref{fig:fourpulses}) is not the one with the best pulse quality.
Despite the different quality of the pulse profiles, we find evidence that the height of the second peak (defined as maximum minus minimum of the corresponding peak of the fitting function) varies significantly compared to that of the first peak. From Jan 22 through Jan 25 the ratio of the two peaks is $0.68 \pm 0.07$, $0.35 \pm 0.08$, $0.56 \pm 0.09$, $0.49 \pm 0.05$, respectively.

\section{Discussion and conclusions}
\label{sect:discussion}

We analyzed the January 2018 Aqueye+ observations of PSR J1023+0038 and detected, for the first time with Aqueye+, a highly significant millisecond pulsation.
The rotational period ($P = 1/\nu_0 = 1.68798744596 \pm 3.9 \times 10^{-10}$ ms) is consistent within the uncertainties with the value extrapolated from the X-ray ephemeris of \cite{2016ApJ...830..122J} and the properties of the pulse shape are similar to those reported by \cite{2017NatAs...1..854A}. Night-to-night variations of the pulse shape are observed, with the height of the second peak varying significantly
(compared to that of the first peak).

The accuracy of our timing and acquisition system allowed us to derive an independent timing solution for PSR J1023+0038 and to measure its absolute rotational phase with high accuracy ($\sim 12$ $\mu$s) using 4 consecutive observing nights. This level of accuracy is needed to search for possible phase shifts between the optical and X-ray pulses, as done for other optical pulsars as the Crab pulsar (e.g. \citealt{2014MNRAS.439.2813Z}) or the Vela pulsar \citep{2019MNRAS.482..175S}. For the Crab pulsar, the optical peak clearly leads the radio peak, but is in phase with the X-ray one (within $\sim 0.003$). A similar evidence for a phase alignment of the optical and X-ray peaks that correspond to the radio peak is found in the Vela pulsar. If the optical and X-ray peaks of PSR J1023+0038 are in phase, this would then suggest that the emission mechanism is the same, likely a rotation powered pulsar. Conversely, finding a significant phase shift bewteen the optical and X-ray peaks would provide independent evidence that a different emission mechanism is at work. For PSR J1023+0038 an accuracy of $\sim 12$ $\mu$s, as that obtained from our timing solution, guarantees to pinpoint a phase shift as small as $\sim 0.01$.

Other optical measurements of the absolute rotational phase of PSR J1023+0038 could be affected by systematic errors related to the drift of the internal clock of the acquisition system. To compensate for this effect, \cite{2017NatAs...1..854A} apply a linear correction to the photon arrival times of their event lists. This linear correction is calibrated comparing the time of arrival of the main peak of the Crab pulsar from an optical observation taken in February 2016 with that reported in the Jodrell Bank monthly ephemeris\footnote{http://www.jb.man.ac.uk/~pulsar/crab.html} \citep{1993MNRAS.265.1003L}. \cite{2017NatAs...1..854A} find that the optical peak leads the radio one by $\sim 180$ $\mu$s, in agreement with previous findings (e.g. \citealt{2012A&A...548A..47G,2014MNRAS.439.2813Z}). However, the Jodrell Bank ephemeris contains monthly averages of the phase measurements and can be affected by errors of $\approx 100$ $\mu$s. For Feb 2016 the reported error is 70 $\mu$s, but weekly phase excursions around the monthly ephemeris could be as large as $\sim 50-100$ $\mu$s, as shown in \cite{2016A&A...587A..99C}. For this reason, considering the accuracy of our timing solution, simultaneous optical measurements of the absolute phase of PSR J1023+0038 with Aqueye+ at Copernicus are desirable. They can serve not only because of their intrinsic scientific value, but also as calibrators for observations taken with other optical facilities.

As far as the X-ray observations are concerned, their timing accuracy depends on the satellite and the instrument. The EPIC-pn instrument onboard {\it XMM-Newton} in the timing mode has an uncertainty of 48 $\mu$s on the absolute timing and a clock drift of $< 10^{-8}$ ss$^{-1}$ \citep{2012A&A...545A.126M}. The clock drift appears to be sufficiently small not to lead to a significant smearing of the pulse profile of PSR J1023+0038 during a 30 ks long observation \citep{2016ApJ...830..122J}. However, the absolute error may lead to a non-negligible uncertainity in the measurement of the phase shift with respect to other measurements at the level of 0.03. Simultaneous Aqueye+ observations could then be used to determine the optical-X-ray shift (if it is $\ga 0.03$) or to cross-calibrate the optical and X-ray measurements.

Unfortunately, because no X-ray observations were performed at the time of the Aqueye+ observations, we cannot measure directly the X-ray-optical time delay using the present dataset. Nor can we select {\it high} and {\it low mode} time intervals to improve the significance of our measurements. However, future simultaneous multiwavelength campaigns in synergy with X-ray telescopes and with SiFAP at TNG will certainly allow us to perform an accurate measurement of the relative shift between the X-ray and optical peaks in the pulse shape of PSR J1023+0038. This type of measurements provide significant constraints to the geometry of the emission region and, ultimately, to the physical mechanism producing the optical and X-ray pulses.

\section*{Acknowledgements}

We thank the referee for the useful comments. We would like to thank M. Barbieri, P. Ochner, L. Lessio, and all the technical staff at the Asiago Cima Ekar Observatory for their valuable operational support. Based on observations collected at the Copernicus telescope (Asiago, Italy) of the INAF-Osservatorio Astronomico di Padova. We acknowledge financial contribution from the grant ASI/INAF n. 2017-14-H.O (projects ``High-Energy observations of Stellar-mass Compact Objects: from CVs to Ultraluminous X-Ray Sources'' and ``Understanding the x-ray variabLe and Transient Sky (ULTraS)''). This research made use also of the following PYTHON packages: MATPLOTLIB \citep{2007CSE.....9...90H}, NUMPY \citep{2011CSE....13b..22V}, and ASTROPY \citep{2013A&A...558A..33A}.







\bsp	
\label{lastpage}
\end{document}